\documentclass{article}
\usepackage[T1]{fontenc} 
\usepackage[utf8]{inputenc} 
\usepackage{ismir,amsmath,cite,url}
\usepackage{graphicx}
\usepackage{color}
\usepackage{booktabs}
\usepackage{microtype}
\usepackage{paralist}
\usepackage{units}
\usepackage{amsfonts}
\usepackage{lineno}

\title{Evaluating Generative Audio Systems and Their Metrics}

\twoauthors
  {Ashvala Vinay} {Center for Music Technology \\ Georgia Institute of Technology}
  {Alexander Lerch} {Center for Music Technology \\ Georgia Institute of Technology}

\begin{document}

\maketitle

\begin{abstract}

Recent years have seen considerable advances in audio synthesis with deep generative models. However, the state-of-the-art is very difficult to quantify; different studies often use different evaluation methodologies and different metrics when reporting results, making a direct comparison to other systems difficult if not impossible. Furthermore, the perceptual relevance and meaning of the reported metrics in most cases unknown, prohibiting any conclusive insights with respect to practical usability and audio quality. This paper presents a study that investigates state-of-the-art approaches side-by-side with (i) a set of previously proposed objective metrics for audio reconstruction, and with (ii) a listening study. The results indicate that currently used objective metrics are insufficient to describe the perceptual quality of current systems.

\end{abstract}

\section{Introduction}\label{sec:introduction}

There has been growing research interest in building deep learning models that are capable of generating audio. Models such as WaveNet \cite{oord_wavenet_2016}, NSynth \cite{engel_neural_2017}, WaveGAN \cite{donahue_adversarial_2019} and, most recently, DDSP \cite{engel_ddsp_2020} paved the way for data-driven Neural Audio Synthesis (NAS). Models like DDSP and NSynth have been advertised in YouTube videos prominently \cite{adam_neely_turning_2020, andrew_huang_music_2017}, where artists make music using neural network generated audio, indicating that there is real world applicability to generative audio models.

Despite the many advances in generative modeling of sounds in the past couple of years, the evaluation of these systems lacks established methodology. Most importantly, systems are not evaluated with a consistent set of metrics. For instance, researchers have suggested the following to evaluate the output of such generative systems: \begin{inparaitem}[]
    \item   the classification accuracies of neural networks trained to classify the sounds into predefined categories such as pitch and sound qualities \cite{engel_neural_2017, nistal_comparing_2021, nistal_vqcpc-gan_2021, donahue_adversarial_2019},
    \item   statistical methods  \cite{kong_diffwave_2021, engel_gansynth_2019}, and
    \item   subjective evaluation through listening studies  \cite{kong_diffwave_2021, donahue_adversarial_2019, engel_gansynth_2019}. 
\end{inparaitem}

This inconsistency in metrics and methodology makes a direct comparison of systems difficult at best, making it hard to understand the current state of the art and to measure the impact of new innovations in the field.

In this study, we measure and compare the quality of the output of neural networks capable of producing short time-invariant samples. The goal is to evaluate state-of-the art systems comparatively with previously used evaluation metrics and to investigate the perceptual relevance of these metrics for measuring audio quality.

To pursue these goals, we trained three widely known neural networks, DDSP \cite{engel_ddsp_2020}, NSynth \cite{engel_neural_2017}, and Diffwave \cite{kong_diffwave_2021}. The evaluation results published with the introduction of these methods all imply that these models synthesize sounds at high quality. However, since different metrics are used for each system, no comparison is possible. To enable such a comparative analysis, we survey and implement a set of metrics and apply them to these systems. Furthermore, we conduct a listening study to measure the perceptual sound quality of the system outputs.

The core contributions of this paper are: 
\begin{inparaenum}[(i)]
    \item a review of currently used metrics for the evaluation of synthesis quality and a comparative analysis of 3 popular neural audio synthesizers,
    \item a listening study for assessing the perceptual audio quality of these synthesizers, and
    \item an investigation on the perceptual relevance of the objective metrics.
\end{inparaenum}  

\section{Evaluation of NAS systems today}\label{sec2}

Generative systems are notoriously difficult to evaluate \cite{theis_note_2016} and new metrics are proposed frequently to understand whether generative networks are able to capture desirable characteristics required for a given task. In audio and music, the task of evaluation is difficult because 
\begin{inparaenum}[(i)]
	\item the ground truth is not well defined, as various outputs might be considered ``correct,'' \cite{caetano_formal_2012} 
	\item the previously used set of objective metrics for NAS most likely misses or insufficiently models perceptual qualities of a sound \cite{ananthabhotla_towards_2019, nistal_comparing_2021},
	\item and aesthetic preferences are, by definition, subjective. \cite{leder_model_2004}
\end{inparaenum}

Some contemporary metrics for NAS derive from literature on evaluating Generative Adversarial Networks (GANs), where diversity of samples and modeling the distribution of data play a significant role \cite{salimans_improved_nodate}. Objective evaluation metrics typically rely on either mathematical formulations of a success measure, or~---in the case of contemporary GAN literature---~using a separate neural network to identify if the model is working appropriately. Accordingly, we categorize the metrics into the four groups 
\begin{inparaenum}[(i)]
    \item   reconstruction metrics, 
    \item 	sample diversity measures,
    \item   distribution distance measures, 
    \item   and measures derived from subjective evaluation methods.
\end{inparaenum}

\noindent
\textbf{Reconstruction errors} are computed as the difference between a given input sound $ S_{i} $ and a generated sound $ S_{g} $. The error is typically defined as either an $ \ell_{2} $ norm (squared error) or a $ \ell_{1} $ norm (absolute error). These differences can be computed on either the time-domain signal or a spectrogram. Typically, the MSE/MAE is computed on spectrograms, given their ubiquity as the input and generated representation for neural networks. MSE and MAE have a range between 0 and infinity, with a MSE/MAE of 0 indicating perfect reconstruction. 

To give examples of practical use, Engel et al. use the multi-scale spectrogram loss, which compares spectrograms across a set of FFT sizes as a measure of reconstruction error and as a loss term for use in training \cite{engel_ddsp_2020}. This was recently used as a metric by Shan et al.\ to compare DDSP with their proposed Differential WaveTable Synthesis \cite{shan_differentiable_2022}.

With respect to reconstruction metrics, there appears to be an ambiguity about the purpose of these metrics, as they seem to be used as both the training loss to be minimized and the evaluation metric itself. Also note that these errors are known to not have a clear perceptual meaning, as a high MSE between two sounds does not necessarily mean that such two sounds will be perceived as dissimilar (or vice versa). There is plenty of evidence in the field of (perceptual) audio coding verifying that measuring the power of the coding error is insufficient to capture the perceptual quality of the sound \cite{thiede_peaq--itu_2000}.

\noindent
\textbf{Sample diversity metrics:}  In GAN literature, a lot of emphasis is placed on the performance of the ``generator'' and to ensure that it is able to produce classifiable samples that capture the diversity of classes from the training dataset. 
The two metrics we will discuss in this section use machine learning and deep learning driven approaches to measure sample diversity. 

\begin{compactitem}
\item
\textbf{Number of statistically Different Bins} (NDB/$k$) is a metric devised to identify mode collapse in GANs, a phenomena where a network produces a lot of outputs that look like alike, therefore lacking sample \textit{ diversity} \cite{richardson_gans_2018}. NDB/$k$ is computed on a Voronoi decomposition from the $ k $-means centroids of the training samples. The clusters are computed directly in the sample space.\footnote{in the case of images, the pixel distance} The $k$ clusters are referred to as the ``bins.'' To compute a score, test samples are assigned to the $k$ clusters/bins using an $L2$ distance measure between the samples and the centroids of the clusters. A two-sample t-test on each bin identifies the statistically different bins. The final NDB score is given by counting the number of statistically different bins and dividing by the number of clusters. 

NDB/$k$ scores are between the range of 0 and 1. According to the interpretation of the score provided by Richardson and Weiss\ \cite{richardson_gans_2018}, a score of 0 indicates that the network is producing a large diversity of samples that captures the training distribution well, whereas a score approaching 1 suggests that the generator has collapsed. 

This was used by Diffwave \cite{kong_diffwave_2021} and GANSynth \cite{engel_gansynth_2019} as part of their evaluation metrics. 

Richardson and Weiss \cite{richardson_gans_2018} state in their definition of the metric that computing the distance for each pixel in an image using an $L2$ distance is perhaps not meaningful. As we have stated above, distances like $L2$ are not good at capturing perceptual qualities of sound, making this of particular concern when evaluating audio and the outputs of generative audio systems.

\item
\textbf{Inception scores (IS)} were proposed by Salimans et al.\ as a way to evaluate image generating GANs by using a classifier \cite{salimans_improved_nodate}. This classifier is used to automatically evaluate whether the output of a GAN was of reasonable quality and captured the \textit{diversity} of samples in the dataset.

The Inception classifier produces class label probabilities for a given input. Ideally, each input produces a high probability for one class label and our generative system is able to produce many of them. At the same time, the generator should be able to produce many such images, that can be classified uniquely into a large set of labels. If the difference between the probability distribution of predicted labels for the generated images and the marginal distribution of the labels from the generated data is small, it implies that the the generator is unable to produce a a diverse number of easily classifiable images. The mathematical formulation of IS can be found in \cite{salimans_improved_nodate}. The score itself has a lower bound of zero and an upper bound of infinity. The higher the score, the better.

Within the context of audio, IS was used in evaluating WaveGAN \cite{donahue_adversarial_2019}, where an Inception network was trained on the SC09 dataset  with spectrogram inputs. More recently, two inception scores have been proposed for NAS: the Pitch Inception Score (PIS) and the Instrument Inception Score (IIS) \cite{nistal_vqcpc-gan_2021}. These measures tell us if generator has captured the true distribution of discrete MIDI pitch classes and the instrument classes in the NSynth dataset \cite{engel_neural_2017}, and are thus focused on inherent sound properties that are not directly related to audio quality.

Salimans et al.\cite{salimans_improved_nodate} stated that the Inception Score for an image generator was found to correlate well with human judgment of image quality \cite{salimans_improved_nodate}. However, no such work has been done in evaluating the perceptual meaningfulness of IS with audio.

\end{compactitem}

\noindent
\textbf{Distribution distance metrics:} Another important facet of evaluating GANs is identifying whether the distribution of data produced by the generator is close to the distribution of real data. Like the Inception score mentioned above, the metrics discussed here use neural networks. The difference here is that these metrics rely on using embeddings from a neural network. 

As noted by Ananthabhotla et al.~\cite{ananthabhotla_towards_2019}, matching distributions cannot guarantee a perceptually closer result.

\noindent
\begin{compactitem}
\item
\textbf{Kernel Inception Distances (KID)} are scores that are generated by computing the distance between embeddings of input and generated data fed to inception networks. The distance is computed using Maximum Mean Discrepancy (MMD), a statistical test that describes the difference between two distributions of data. They were first introduced in a paper by Binkowski et al.~\cite{binkowski_demystifying_2018} where they used MMD to train a critic or discriminator and KID was shown as a metric to evaluate the convergence of the GAN. In order to compute KID, we compute a distribution of embeddings extracted from the Inception classifier for both the reference and generated output and compute the MMD between the distributions of reference and generated embeddings. The score is defined with a lower-bound of zero and an upper bound of infinity.

This was used by Nistal et al.\ to compare input feature representations using GANs \cite{nistal_comparing_2021} and in a separate paper by Nistal et al.\ to measure the performance of an architecture they built called the VQCPC-GAN \cite{nistal_vqcpc-gan_2021}. It was also used to evaluate DarkGAN \cite{nistal_darkgan_2021}. 

\item
\noindent
\textbf{Fréchet Audio Distance (FAD)} \cite{kilgour_frechet_2019} is a metric originally developed to evaluate sound enhancement algorithms, but recently it has found use in evaluating NAS systems. 
The computation of the FAD relies on the VGGish embeddings for both the reference and the generated sounds. The VGGish embeddings are then fitted to multi-variate gaussians. The FAD itself is the Fréchet distance between the two distributions $\mathcal{N}_{r}$ and $\mathcal{N}_{g}$ representing the reference and generated gaussian distributions. The mathematical definition can be found in \cite{kilgour_frechet_2019}.

This metric was used for evaluating Diffwave \cite{kong_diffwave_2021}, Neural Waveshaping Synthesis \cite{hayes_neural_2021}, DarkGAN \cite{nistal_darkgan_2021}, and CRASH \cite{rouard_crash_2021}. 

\end{compactitem}

\subsection{Subjective evaluation}

Since the quality of the generated outputs is ultimately a perceptual property, listening studies have been previously used to evaluate a neural network's generated audio quality.  A lot of listening studies use a popular method called ``Mean Opinion Score'' (MOS) \cite{ribeiro_crowdmos_2011}.  Users participating in the listening study are asked to rate the sound they hear on a Likert scale between 1 and 5 across a set of questions. For example, WaveGAN asked participants to rate sounds on their ``sound quality, ease of intelligibility, and
speaker diversity''  \cite{donahue_adversarial_2019} where 1 indicates  bad and 5 indicates excellent. These questions are asked for a collection of methods that the researcher seeks to evaluate. 

The MOS survey strategy has been used for the evaluation of WaveGAN \cite{donahue_adversarial_2019}, Diffwave \cite{kong_diffwave_2021}, and in neural speech generation literature \cite{prenger_waveglow_2019, wang_tacotron_2017}. Kong et al.\ compared their results to WaveGAN using MOS and found that their network scored higher \cite{kong_diffwave_2021}.

It should be noted that surveys that use MOS do not explicitly ask participants to compare the outputs against a reference and instead present participants with a single sound for every question. This means that MOS surveys give you an absolute rating for every sound, not a relative preference. The absence of reference precludes the ability to rank the methods that are being evaluated. For example, a person might like sound A and sound B in isolation and provide high ratings to both, however, it remains unclear whether the person has a relative preference for sound A or B.

A well known alternative to MOS surveys is to use a survey method called MUltiple Stimuli, Hidden Reference and Anchor (MUSHRA) \cite{international_telecommunications_union_last_bs1534_nodate} . It is an ITU recommendation for studies designed to evaluate differences in audio quality between audio codecs \cite{zielinski_potential_2007}. While it hasn't seen significant usage in evaluating NAS, it was recently used by Hayes et al. \cite{hayes_neural_2021} in evaluating the performance of their Neural Waveshaping Synthesis (NeWT) model against the performance of DDSP. Their listening study results showed that their neural network outperformed DDSP on most instruments, with the exception of violins. This was shown to correlate well with the computed FAD scores for DDSP and NeWT. 

\begin{table*}
\small
\centering
\begin{tabular*}{\textwidth}{l @{\extracolsep{\fill}} |c|c|c|c|c|c|c|c}
{System}\phantom{......} &  NDB/k ($\downarrow$) &    PKID($\downarrow$) &    IKID($\downarrow$) &     PIS($\uparrow$) &     IIS($\uparrow$)	  &   MSE($\downarrow$)   &             MAE ($\downarrow$) & FAD ($\downarrow$)\\
\hline\hline
Diffwave &   0.74 &  0.0093 &  0.0021 &  2.3814 &  \textbf{5.6477} &  0.0291 & 0.1369 & 7.9488\\
\hline
{DDSP}    &   \textbf{0.20} &  \textbf{0.0053 }&  0.0020 &  \textbf{3.3224} &  5.3371 & \textbf{0.0130} & \textbf{0.0666} & \textbf{1.1519}\\
\hline
NSynth   &   0.74 &  0.0101 &  0.0024 &  2.3238 &  4.6364 & 0.0329 & 0.1224 & 4.0590\\
\hline
Anchor & 0.72 & 0.0123 & \textbf{0.0006} & 2.9356 & 5.3017 & 0.0257 & 0.0857 & 1.4952\\
\end{tabular*}
\caption{Table with objective results for each of the neural networks that we measured. $\downarrow$ indicates that a lower score is better and $\uparrow$ indicates that a higher score is considered better. Bold indicates best performance.} 
\label{table:res_1}
\end{table*}

\section{Experimental setup} 
Aiming at our goal of comparing the output quality of popular generative systems and assessing the metrics commonly used for evaluation, we chose three neural networks and re-trained DiffWave and DDSP with the NSynth dataset and used the NSynth network directly. We report both objective metrics and subjective ratings of the outputs of these system.

We break our study into the three phases:
\begin{inparaenum}[(i)]
    \item comparative analysis using objective metrics, 
    \item comparative analysis using listening study results, and 
    \item a brief investigation into the perceptual relevance of the objective metrics.
\end{inparaenum}

\subsection{Dataset}

The NSynth dataset is a publicly available dataset with approximately 300k sounds of \unit[4]{s} length spanning acoustic and electronic timbres \cite{engel_neural_2017}. The sounds are all sampled at \unit[16]{kHz}. It is comprised of 11 instrument families. There are ten unique ``quality'' descriptors that are attached to every sound in the dataset. The descriptors are timbral, for e.g ``dark,'' ``bright,'' ``reverb,'' and ``percussive.'' The NSynth dataset is frequently used in generative audio research and is therefore an obvious choice for this study. 

The NSynth dataset's test set was used to generate all the samples that were used in both the listening study and to compute objective metrics.

\subsection{Models}

Model selection was based on the criteria of age, architecture, and public availability. There are a number of generative audio systems that have been published since 2017, but not all of them were available publicly or were difficult to train due to the lack of computational resources. Models selected were NSynth \cite{engel_neural_2017}, DDSP\cite{engel_ddsp_2020}, and DiffWave\cite{kong_diffwave_2021}. These models are widely known and represent different architectural designs. 

\textit{NSynth} \cite{engel_neural_2017} is a deep learning based generative audio system that uses a Variational Autoencoder architecture that learns to generate musical instrument timbres with the ability to be controllable. Its primary novelty is the fact that it can interpolate between multiple sounds and generate new timbres in the process. NSynth is the ``oldest'' of the evaluated systems. The publicly available weights for the NSynth model were used for this study to generate the sounds. 

\textit{DDSP} \cite{engel_ddsp_2020} uses classical synthesis techniques like additive synthesis and noise filtering in the context of deep learning by treating them as differentiable blocks.  It uses an encoder-decoder architecture that produces the fundamental frequency, loudness envelope and the necessary variables to control the additive synthesizer. DDSP is also known for it ability to perform ``timbre-transfer,'' where it takes sounds produced from one instrument and outputs it on a different instrument. To train DDSP, we used publicly available code for training with the NSynth dataset and verified it worked by producing metrics similar to the metrics reported in the original paper.

\textit{Diffwave} \cite{kong_diffwave_2021} is the most recent system that uses a new generative modeling technique called ``Diffusion'' on audio. Diffusion models start with white noise and through a fixed number of iterations, learn to generate audio. During training, the models use a forward and backward process, where the forward process takes the reference, corrupts it with white noise iteratively until the whole signal is noise. The backwards process learns how to iteratively remove the white noise to recover the reference.  Diffwave was primarily evaluated on speech and produced results that seemed to outperform WaveGAN and Wavenet significantly.

DiffWave was trained with an implementation available online\footnote{https://github.com/lmnt\-com/diffwave}. Since this algorithm was not originally trained with NSynth, we had to verify if our model worked properly. We trained the network for 2 million steps and evaluated the network's automatic metrics and found that the metrics aligned with the paper's reported results.

\subsection{Objective metrics}

To evaluate our neural networks, we used the set of metrics described above. As we discussed in Section \ref{sec2}, some of the objective metrics were designed with specific architectures like GANs in mind. However, since the metrics have found use in the evaluation of other types of architectures we chose to evaluate all of our networks with the same set of metrics.

The pre-trained pitch and instrument inception networks trained on the NSynth dataset from Nistal et al.'s paper on comparing audio representations \cite{nistal_comparing_2021} were used to compute neural network driven metrics such as as PKID and IIS and the ``official'' implementations of 
NDB/$k$\footnote{https://github.com/eitanrich/gans\-n\-gmms} and FAD\footnote{https://github.com/google\-research/google\-research/} were used. In order to compute NDB/$k$, we trained the K-means clustering on the NSynth dataset.

The rest of the metrics, like MSE and MAE were computed using SciKit-learn's built-in metrics \cite{pedregosa_scikit-learn_2011}. We tried to include other metrics but could not select them due to high variance in scoring between implementations (such as PEAQ \cite{thiede_peaq--itu_2000}).

To summarize, the following objective metrics are computed for the network outputs:

\begin{inparaenum}
	\item NDB/$k$,
	\item PKID,
	\item IKID,
	\item PIS,
	\item IIS,
	\item MSE,
	\item MAE, and
	\item FAD.
\end{inparaenum}

\subsection{Listening study}\label{list-spec}

The listening study uses a variation of the aforementioned MUSHRA methodology. It is popularly used in evaluating ``intermediate'' differences in low bitrate audio codecs \cite{zielinski_potential_2007}. Given that we are measuring audio reconstruction, we believe that treating the neural network outputs similar to encoded audio is a suitable choice for evaluating audio quality differences. The rating scale used in the study is divided according to the MUSHRA specification. A score between 0 and 20 indicates that the sound is rated \textit{bad}, 20 to 40 indicates that the sound is rated \textit{poor}, 40 to 60 is considered \textit{fair}, 60 to 80 is considered \textit{good}, and 80 to 100 is considered \textit{excellent}.

The 4096 sounds from the NSynth test set are reconstructed using the three generative systems. In addition, one anchor sound is generated for each test sample by low pass filtering the sound at a \unit[1]{kHz} cutoff frequency and reducing its bit depth to \unit[8]{Bits}.

Participants were recruited by emails sent to two large academic audio-focused communities. Participants were asked to use a good pair of headphones or speakers in order to participate in the survey. Prior to the listening study, the participants were asked to share their age bracket, experience with audio synthesis, and how much money they have spent on the audio equipment they used. This then led to a training phase where participants were presented with an example sound and necessary introduction to the survey.

When presenting the survey, a sound is randomly selected out of 3237 possible sounds with MIDI note numbers ranging from 22 to 84.\footnote{We used the note number range to remove sounds that were either inaudible or could potentially be uncomfortable to listen to.} Five sliders are presented in random order, with three sliders referring to the generated audio output of the three neural networks to evaluate, and the other two sliders corresponding to the hidden reference and the anchor. The participants were asked to rate audio generated by each neural network on its audio quality compared to the reference. This presentation is repeated ten times without collision (i.e., no two sounds are ever repeated for a participant). 

To clean up the data, responses that are ``incomplete'' are removed entirely from data that can be used for analysis. MUSHRA analysis generally removes raters who rate the reference below a high rating threshold \cite{mendonca_statistical_2018}; thus, participants who rate the reference below an average of $85$ will be removed. 

Statistical analysis of MUSHRA and MUSHRA-like data relies on running statistical tests such as ANOVA or Wilcoxon tests \cite{mendonca_statistical_2018} to measure differences in results between the presented conditions. We will use the Wilcoxon tests for the ratings to measure statistical differences between presented conditions. The test indicates differences between the generative models. The MUSHRA results will also be broken down by demographic information collected above to measure if different demographic groupings rated the models differently.

In order to compare objective and subjective ratings, we will also compute rankings for both ratings. We are interested in knowing how frequently specific networks were considered the best and comparing it to the rankings for each metric. Rank driven correlations were used by Ycart et al.\ \cite{ycart_investigating_2020} in their paper validating perceptual metrics related to piano transcription.

\section{Results} 

\subsection{Objective metric results} 

The results from the objective metrics as discussed above are shown in Table~\ref{table:res_1}. Overall, these results seem to indicate that DDSP is producing higher quality samples that are more easily classifiable compared to DiffWave and NSynth.

 \begin{figure}
  \centerline{
  \includegraphics[width=0.9\columnwidth]{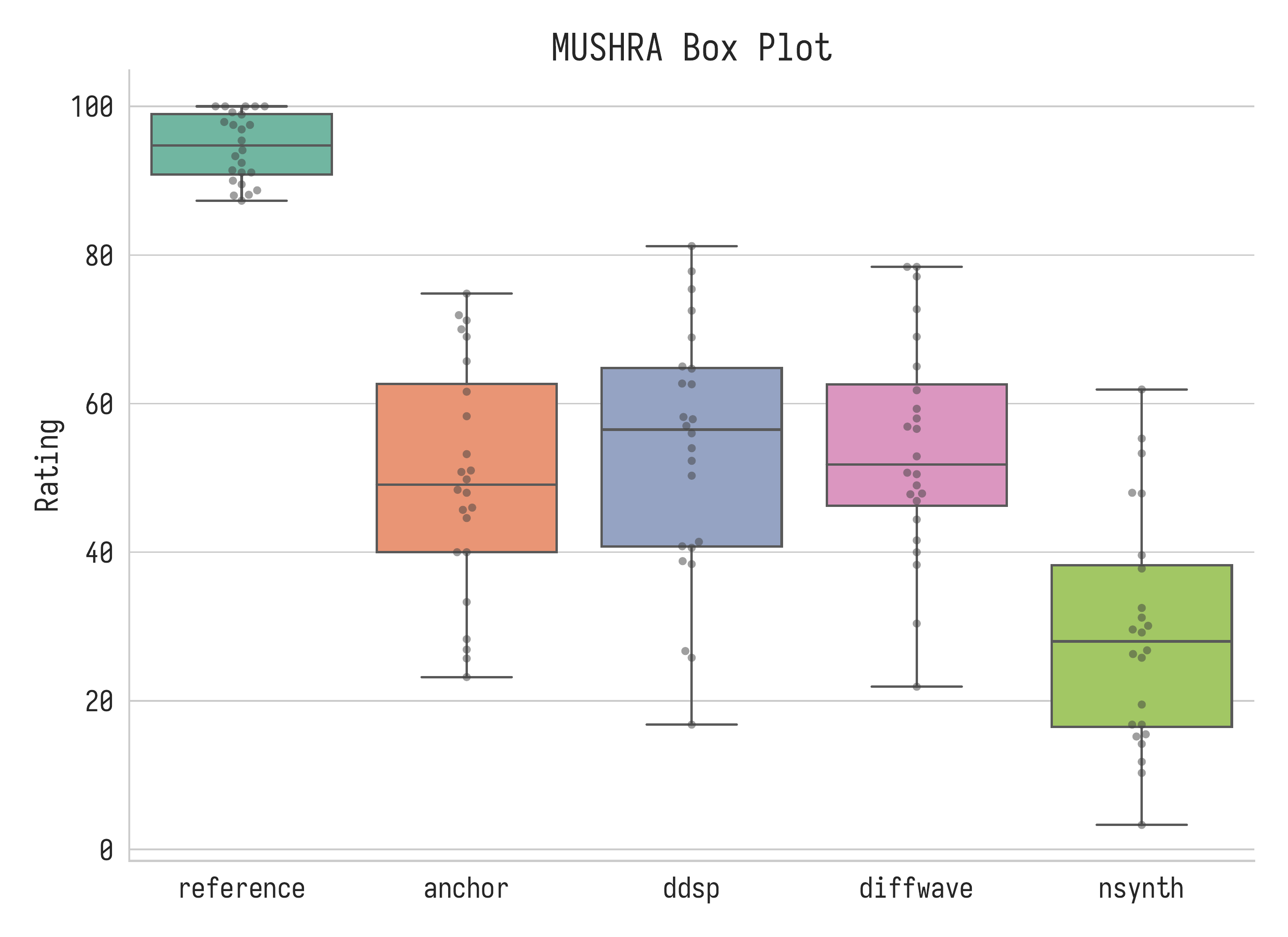}}
  \caption{A boxplot of ratings from participants in the listening study. The size of the box represents the interquartile range of ratings from the listeners.}

  \label{fig:mushra}
 \end{figure}
 
 DDSP has a 54\% better PKID score over Diffwave and a 62\% higher score than NSynth. This trend continues across most metrics, with the exceptions of the IIS, where Diffwave achieves a 5\% higher score than DDSP, and the IKID where there is only a 4\% difference between DDSP and Diffwave. The kernel distances mentioned here do not measure audio quality.

The MSE and MAE results tell us that DDSP is the ``closest'' to the reference sounds, with an MSE of 0.0130, a 76\% and 79\% improvement over Diffwave and NSynth, respectively.

DDSP significantly outperforms Diffwave and NSynth on the Fréchet Audio Distance. FAD is also a category where NSynth outperforms Diffwave. This metric indicates that~---according to the VGGish representation---~DDSP and NSynth are generating samples that are closer to the reference than Diffwave. 

Based on these metrics, we can state that DDSP outperforms Diffwave and NSynth and produces a diverse set of easily classifiable samples and produces samples that are much closer to the reference sounds.

\subsection{Listening study results} 

Data was collected from 77 participants from our listening study. After data cleanup, a total of 24 submissions were considered trustworthy.  74\% of our participants were between the age of 24 and 50, 22\% of our participants between 18 and 24 and 3.7\% or 1 participant over the age of 50. 37\% of our participants reported that they were very familiar with music production tools and 18.5\% reporting that they were extremely familiar. 29\% of the participants reported that they were moderately knowledgeable about sound synthesis techniques, with an even distribution of familiarity ranging from ``Not very knowledgeable'' to ``Extremely knowledgeable.'' 33\% of our raters reported that they had spent over \$750 on their audio equipment and 25\% having spent between \$250 and \$500.

The overall visualization of the responses can be seen in Fig.~\ref{fig:mushra}. Every dot in the chart is an averaged rating from a participant. As expected, the reference scored highly with an average rating of $92$. The analysis of the listening study results shows that DDSP and DiffWave scored similarly while NSynth performed considerably worse than the other networks. Both DDSP and Diffwave have average ratings around $53$ while   sounds generated by NSynth received an average rating of approx.~$29$.

The inter-rater Krippendorff $\alpha$ score\cite{krippendorff_computing_2011} is 0.66\footnote{Krippendorff $\alpha$ ranges from -1 to 1, where -1 indicates significant disagreement between raters and 1 suggests maximal agreement}, suggesting that the subjects were largely in agreement. 

The Wilcoxon test for statistical significance was applied to the data \cite{mendonca_statistical_2018}. We found no statistically significant difference between DDSP/Diffwave ($p=0.629$) and a statistically significant difference between DDSP/NSynth ($p=  8 \times 10^{-6}$) and Diffwave/NSynth ($p = 8 \times 10^{-6}$). There was a significant difference between the Reference and all the systems and the anchor ($p\leq 8 \times 10^{-6} $). There is a significant difference between Anchor/NSynth ($p = 8 \times 10^{-6} $), but no statistically significant difference between Anchor/Diffwave or Anchor/DDSP, with p-values of $ 0.144 $ and $ 0.4385 $, respectively. 

Breaking down the results by the subjects' self-reported familiarity with audio synthesis technologies, we first investigate the inter-rater variation within different ``expert levels.'' Participants who said they were the least knowledgeable had an Krippendorff $\alpha$ of $0.7$, moderately knowledgeable raters had a Krippendorff $\alpha$ of $0.608$ while participants who were the most knowledgeable had a Krippendorff $\alpha$ of $0.9$. 
Participants who reported to be either very or extremely knowledgeable tended to rate DDSP and Diffwave lower than the overall scores (DDSP: $47$, Diffwave: $43$). Less knowledgeable participants tended to rate Diffwave and DDSP higher than the overall scores (DDSP: $56$, Diffwave: $59$). There was a statistically significant difference in how these two groups rated Diffwave ($p=0.006$) but no statistically significant difference in how they rated DDSP, NSynth, or the reference and anchor point. 

We found no statistically significant differences in ratings between the age categories or in ratings based on money spent on audio equipment.  

\noindent
\textbf{Instrument results:}
To investigate whether specific instruments or instrument groups are consistently rated higher or lower than others, the listener ratings were broken down into ratings by instrument family (according to the NSynth dataset). We found no statistically significant differences in ratings between DDSP and Diffwave, but found statistically significant differences between DDSP/NSynth and Diffwave/NSynth. Additionally, there were no statistically significant differences between instrument sources and the ratings from the listeners. 

The IKID metric tell us that all three networks are producing samples that are  close to the reference and the IIS score tells us that Diffwave should be slightly better than DDSP and much better than NSynth. While we cannot state that this result is accurate for IKID, it is in line with the results from the IIS computation.

\noindent
\textbf{Comparing the listener ratings to objective metrics: } In order to identify if our listening study results lined up with our objective metrics, we took the selection of sounds that were presented to participants and computed the correlation between their sample based metrics and our ratings using the Pearson correlation coefficient, which tells us how correlated two samples are from a range of -1 to 1. We have sample based results for MSE and MAE, since all the other metrics in our list rely on computing a difference across a distribution of samples. We also computed the Spearman R score and the $R^2$ from a linear regression between the ratings and the MSE and MAE. We found that there is no statistically significant correlation between the MSE/MAE errors and the listener ratings. 

\begin{figure}
  \centerline{
  \includegraphics[width=0.9\columnwidth]{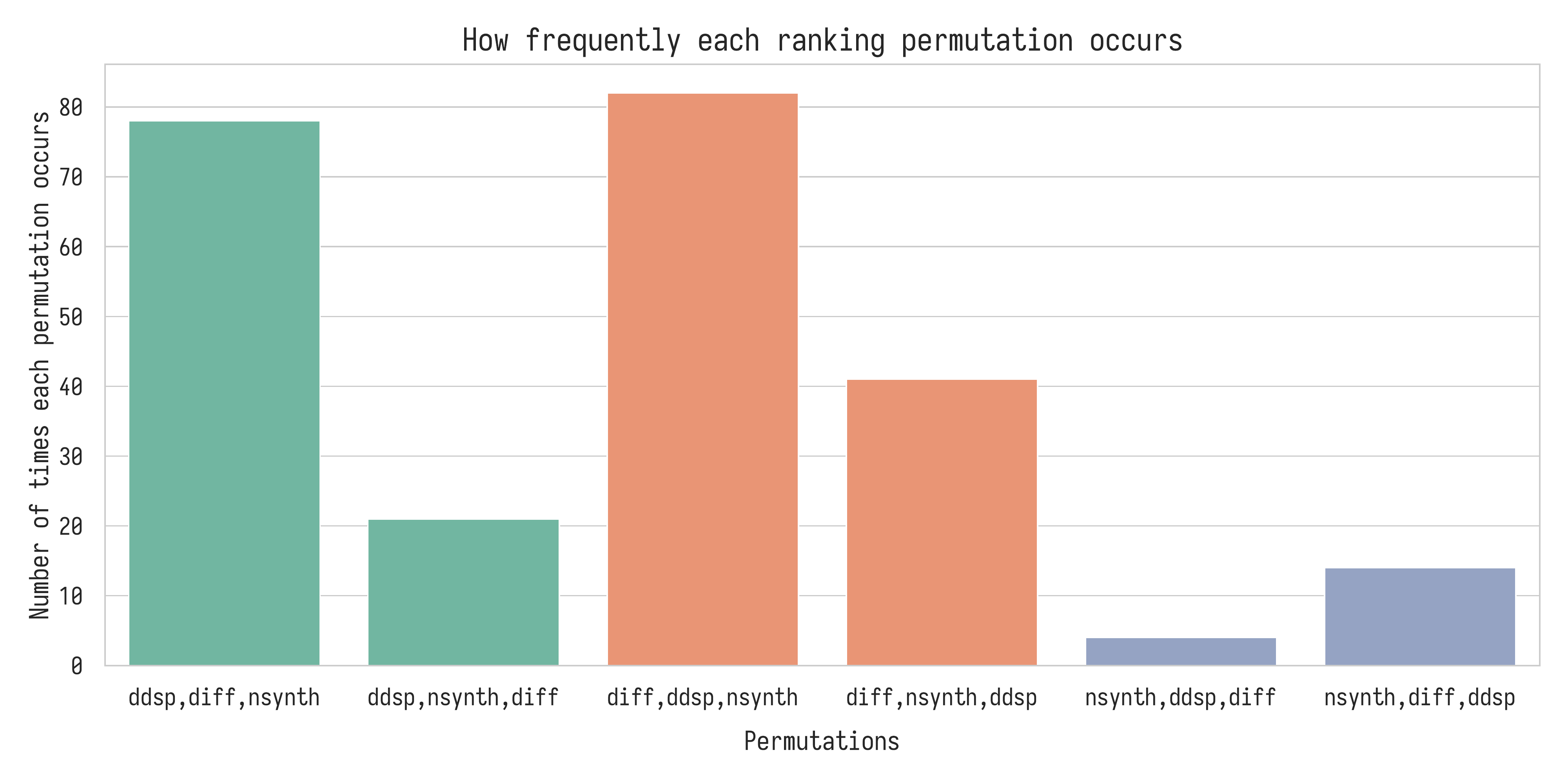}}
  \caption{The plot here describes how frequently a unique permutation of rankings occurred. Diffwave is abbreivated to ``diff''. }
  \label{fig:rankings}
 \end{figure}

The listening responses were also ranked based on how frequently a specific network was ranked higher than the others. In the 240 responses, the listeners rated Diffwave higher than DDSP slightly more frequently (123 vs.\ 99). NSynth was rated the lowest most frequently (163 times). The ranking breakdown showing how frequently each of the 6 permutation of rankings between the 3 networks is shown in \ref{fig:rankings}. We ran a Wilcoxon test on pairings rankings of each network by the listeners and found that the rankings were different with high statistical significance ($p$ < 0.05). 

When ranking the objective metrics, it is clear that DDSP is ranked the best because it has the best results in seven out of the eight metrics, with Diffwave and NSynth in second and third place. However, the rankings from our listening study tell us that the Diffwave is usually the preferred network. This tells us that the metrics are perhaps not entirely reflective of the perceptual audio quality of the networks.

\section{Conclusion} 
We presented a systematic evaluation of three popular systems for neural audio synthesis, comparing them with a set of previously used objective metrics as well as a listening study. The results clearly show that
\begin{inparaenum}[(i)]
    \item   no previously used objective metric captures the perceptual quality of the synthesized sounds sufficiently well,
    \item   any quality ranking based on these objective metrics is questionable, 
    \item   of the three evaluated audio generators, there is no clear listener preference between DDSP and Diffwave, but NSynth is rated lowly, and
    \item  the subjects rate Nsynth worse than the \unit[8]{Bit} anchor but rate DDSP and Diffwave similar to the anchor indicating that there is still considerable work to do on the quality of neural audio synthesis
\end{inparaenum}

These results should give pause to research in the field of neural audio synthesis. How can progress with respect to the audio quality be measured if all available metrics are unable to provide meaningful estimates of audio quality. Many of the objective metrics that we discussed in this paper were designed with the intent of measuring the performance of the sound generating component of the networks, i.e., can the generator produce
\begin{inparaenum}[(i)] 
	\item a diverse set of sounds,
	\item a distribution of samples that are close to the target samples in a relevant dimension, and
	\item accurate reconstructions.
\end{inparaenum}

Our results indicate that \textit{measuring generator performance is insufficient to measure audio quality}. We believe that research in this space should not only include subjective results, it should also include greater efforts into critically evaluating the audio quality of network outputs with meaningful objective metrics.

In future work, we will investigate whether other objective metrics might be more meaningful than the ones evaluated here, or will start a research project on developing a more meaningful quality measure for audio synthesis.

\bibliography{ISMIR}

\end{document}